\documentclass[english, twocolumn, 10pt, aps, superscriptaddress, floatfix, showpacs, prb, citeautoscript]{revtex4-1}
\pdfoutput=1
\usepackage[utf8]{inputenc}
\usepackage[T1]{fontenc}
\usepackage{verbatim}
\usepackage{units}
\usepackage{mathtools}
\usepackage{amsmath}
\usepackage{amssymb}
\usepackage{graphicx}
\usepackage{wasysym}
\usepackage{siunitx}
\usepackage{bm}
\usepackage{xcolor}
\usepackage[colorlinks, citecolor={blue!50!black}, urlcolor={blue!50!black}, linkcolor={red!50!black}]{hyperref}
\usepackage{bookmark}
\usepackage{tabularx}
\usepackage{microtype}
\usepackage{babel}
\usepackage{cleveref}
\usepackage{soul}
\hypersetup{pdfauthor={Put authors},pdftitle={Supercurrent Interference in Few-Mode Nanowire Josephson Junctions}} 
\setstcolor{blue}
\begin{document}

\title{Supercurrent Interference in Few-Mode Nanowire Josephson Junctions}

\author{Kun Zuo}
\thanks{These authors contributed equally}
\affiliation{QuTech, Delft University of Technology, 2600 GA Delft, The Netherlands}
\affiliation{Kavli Institute of Nanoscience, Delft University of Technology, 2600 GA Delft, The Netherlands}

\author{Vincent Mourik}
\thanks{These authors contributed equally}
\affiliation{QuTech, Delft University of Technology, 2600 GA Delft, The Netherlands}
\affiliation{Kavli Institute of Nanoscience, Delft University of Technology, 2600 GA Delft, The Netherlands}
\affiliation{Centre for Quantum Computation and Communication Technologies, School of
Electrical Engineering and Telecommunications, UNSW Sydney, Sydney, New
South Wales 2052, Australia}

\author{Daniel B. Szombati}
\affiliation{QuTech, Delft University of Technology, 2600 GA Delft, The Netherlands}
\affiliation{Kavli Institute of Nanoscience, Delft University of Technology, 2600 GA Delft, The Netherlands}
\affiliation{Australian Research Council Centre of Excellence for Engineered Quantum Systems, St Lucia, Queensland 4072, Australia}
\affiliation{School of Mathematics and Physics, The University of Queensland, St Lucia, Queensland 4072, Australia}

\author{Bas Nijholt}
\affiliation{Kavli Institute of Nanoscience, Delft University of Technology, 2600 GA Delft, The Netherlands}

\author{David J. van Woerkom}
\affiliation{QuTech, Delft University of Technology, 2600 GA Delft, The Netherlands}
\affiliation{Kavli Institute of Nanoscience, Delft University of Technology, 2600 GA Delft, The Netherlands}
\affiliation{Department of Physics, ETH Zurich, CH-8093 Zurich, Switzerland}

\author{Attila Geresdi}
\affiliation{QuTech, Delft University of Technology, 2600 GA Delft, The Netherlands}
\affiliation{Kavli Institute of Nanoscience, Delft University of Technology, 2600 GA Delft, The Netherlands}

\author{Jun Chen}
\affiliation{Department of Physics and Astronomy, University of Pittsburgh, Pittsburgh, PA 15260, USA}

\author{Viacheslav P. Ostroukh}
\affiliation{
Instituut-Lorentz, Universiteit Leiden, P.O. Box 9506, 2300 RA Leiden, The Netherlands}

\author{Anton R. Akhmerov}
\affiliation{Kavli Institute of Nanoscience, Delft University of Technology, 2600 GA Delft, The Netherlands}

\author{Sebasti\'{e}n R. Plissard}
\affiliation{Kavli Institute of Nanoscience, Delft University of Technology, 2600 GA Delft, The Netherlands}
\affiliation{Department of Applied Physics, Eindhoven University of Technology, 5600 MB Eindhoven, The Netherlands}

\author{Diana Car}
\affiliation{QuTech, Delft University of Technology, 2600 GA Delft, The Netherlands}
\affiliation{Kavli Institute of Nanoscience, Delft University of Technology, 2600 GA Delft, The Netherlands}
\affiliation{Department of Applied Physics, Eindhoven University of Technology, 5600 MB Eindhoven, The Netherlands}

\author{Erik P. A. M. Bakkers}
\affiliation{QuTech, Delft University of Technology, 2600 GA Delft, The Netherlands}
\affiliation{Kavli Institute of Nanoscience, Delft University of Technology, 2600 GA Delft, The Netherlands}
\affiliation{Department of Applied Physics, Eindhoven University of Technology, 5600 MB Eindhoven, The Netherlands}

\author{Dmitry I. Pikulin}
\affiliation{Station Q, Microsoft Research, Santa Barbara, California 93106-6105, USA}
\affiliation{Department of Physics and Astronomy, University of British Columbia, Vancouver BC, Canada V6T 1Z1}
\affiliation{Quantum Matter Institute, University of British Columbia, Vancouver BC, Canada V6T 1Z4}

\author{Leo P. Kouwenhoven}
\affiliation{QuTech, Delft University of Technology, 2600 GA Delft, The Netherlands}
\affiliation{Kavli Institute of Nanoscience, Delft University of Technology, 2600 GA Delft, The Netherlands}
\affiliation{Station Q Delft, Microsoft Research, 2600 GA, Delft, The Netherlands}

\author{Sergey M. Frolov}
\affiliation{Kavli Institute of Nanoscience, Delft University of Technology, 2600 GA Delft, The Netherlands}
\affiliation{Department of Physics and Astronomy, University of Pittsburgh, Pittsburgh, PA 15260, USA}

\date{\today}

\begin{abstract}
Junctions created by coupling two superconductors via a semiconductor nanowire in the presence of high magnetic fields are the basis for the potential detection, fusion and braiding of Majorana bound states.
We study NbTiN/InSb nanowire/NbTiN Josephson junctions and find that the dependence of the critical current on the magnetic field exhibits gate-tunable nodes. 
This is in contrast with a well-known Fraunhofer effect, under which critical current nodes form a regular pattern with a period fixed by the junction area.
Based on a realistic numerical model we conclude that the Zeeman effect induced by the magnetic field and the spin-orbit interaction in the nanowire are insufficient to explain the observed evolution of the Josephson effect. 
We find the interference between the few occupied one-dimensional modes in the nanowire to be the dominant mechanism responsible for the critical current behavior. 
We also report a strong  suppression of critical currents at finite magnetic fields that should be taken into account when designing circuits based on Majorana bound states. \end{abstract}

\maketitle

Semiconductor nanowires coupled to superconductors form a promising platform for generating and investigating Majorana bound states \cite{kitaev2001unpaired, oreg2010helical, lutchyn2010majorana,mourik2012signatures, deng2016majorana, albrecht2016exponential,chen2016phasediagram}. 
Josephson weak links based on nanowires may provide additional evidence for Majorana bound states, e.g. through the fractional Josephson effect \cite{molenkamp2016abs4pi,bocquillon2016gapless,Deacon2016josephsonradiation}. 
These weak links can also become elements of Majorana-based topological quantum circuits \cite{Hyart2013braiding, Aasen2016briading, Karzig2016braiding, plugge2016majorana}.
Previous work on semiconductor nanowire Josephson junctions demonstrated supercurrent transistors \cite{doh2005tunable}, transport through few channels \cite{goffman2017conduction}, a nonsinusoidal current-phase relationship \cite{spanton2017current}, nanowire superconducting quantum interference devices (SQUIDs) \cite{vanDam2006supercurrent, szombati2015josephson}, and gate-tunable superconducting quantum bits\cite{deLange2015hyrbidnanowire,marcus2015hyrbidnanowire}.
Recent works reported Josephson effects at high magnetic fields, sufficient to generate unpaired Majorana bound states \cite{szombati2015josephson,giazaotto2015pbinas,giazotto2016magnetically,gharavi2016nb}.

In this Letter we study the critical current as a function of the magnetic field and gate voltage in nanowire Josephson junctions tuned to the mesoscopic few-mode regime.
The junctions consist of InSb weak links and NbTiN superconductor contacts.
For magnetic fields parallel to the nanowire, we observe a strong suppression of the critical current at magnetic fields on the scale of $\SI{100}{mT}$. 
When the magnetic field exceeds $\sim \SI{100}{mT}$, the critical current exhibits aperiodic local minima (nodes). 
In contrast with supercurrent diffraction in large multimode junctions, the magnetic field nodes of the critical current are strongly tunable by the voltages on local electrostatic gates, and are not uniquely determined by the junction geometry and supercurrent density distribution. 
To understand our data, we develop a numerical model of a quasiballistic few-mode nanowire of realistic geometry.
Our model includes the intrinsic spin-orbit effect, as well as the vector-potential and Zeeman effects of the external magnetic fields. 
Based on the simulations, we conclude that quantum interference between supercurrents carried by different transverse modes is the dominant mechanism responsible for both the critical current suppression, and the gate-sensitive nodes in the critical current.

\begin{figure}[ht]
\includegraphics[width=\columnwidth]{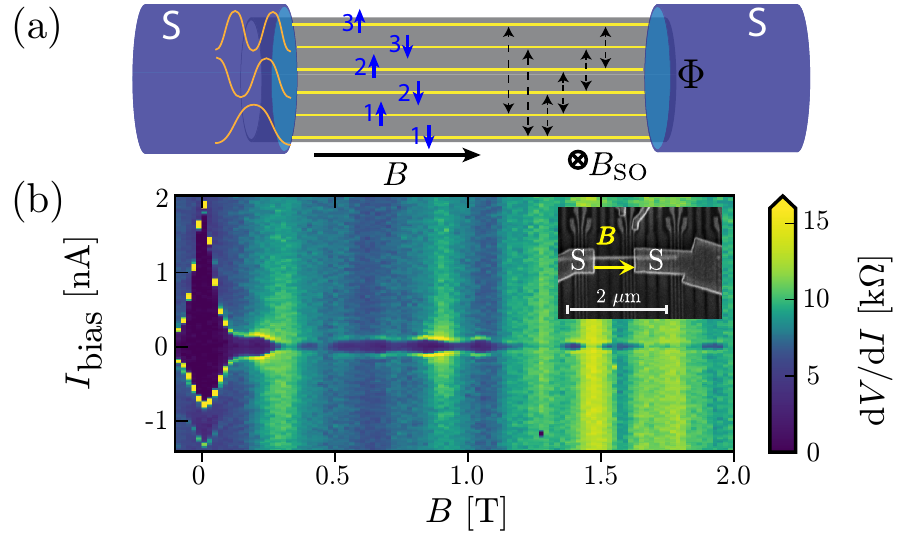}
\caption{(a) Schematic superconductor ($S$)-nanowire-$S$ Josephson junction. The cross section shows cartoon wave functions of $n=3$ transverse modes and the flux $\Phi$ penetrating the area of the nanowire. The blue arrows indicate spin-resolved modes; the black dashed arrows are same-spin scattering events within the wire. All modes are coupled at the contacts. 
The directions of $B$ and the spin-orbit effective field $B_\mathrm{SO}$ are indicated. (b) Differential resistance $\mathrm{d}V/\mathrm{d}I$ versus $B$ and $I_\mathrm{bias}$.
The current bias sweep direction is from negative to positive. Data from device 1. Inset: SEM image of a typical device similar to those studied here. $S$ labels the superconducting contacts while $B$ indicates the in-plane magnetic field for device 2.}
\label{fig:figure1}
\end{figure}

Figure \ref{fig:figure1}(a) presents a schematic of a few-mode nanowire Josephson junction.
The inset of Fig.~\ref{fig:figure1}(b) shows a device similar to those used in this study and their fabrication process is described in Ref.~\onlinecite{mourik2012signatures}.
The junction consists of an InSb nanowire with a diameter of $100 \pm \SI{10}{nm}$ with 80 nm thick dc magnetron sputtered NbTiN contacts.
The wire sits on top of an array of 50 or $\SI{200}{nm}$ wide gates isolated from the junction by a dielectric. 
We report data from devices 1 and 2 in the main text and show additional data from device 3 in the Supplemental Material, Ref. ~\onlinecite{supp}.
Device 1(2) has a contact spacing of $\sim \SI{1}{\micro \m}$($\sim \SI{625}{nm}$) and the nanowire is at an angle of $25^\circ \pm 5^\circ$($0^\circ \pm 5^\circ$) with respect to $B$. 
Device 3 has a shorter contact spacing of $\sim \SI{150}{nm}$ and shows similar behavior of gate-tunable nodes but the initial critical current decay is extended to 400 mT.
The measurements were performed in a dilution refrigerator with a base temperature of $\sim\SI{60}{mK}$. 
All bias and measurement lines connected to the device are equipped with standard $RC$ and copper powder filtering at the mixing chamber stage to ensure a low electrical noise environment. 
The voltage measurements are performed in the four-terminal geometry.

We set all the gates underneath the nanowire to positive voltages, in the few-mode transparent regime in which no quantum dots are formed between the superconducting contacts, and the normal state conductance exceeds $2e^2/h$ (see the full gate trace of the supercurrent in the Supplemental Material ~\cite{supp}).

Figure \ref{fig:figure1}(b) shows a typical example of the differential resistance $\mathrm{d}V/\mathrm{d}I$ as a function of the magnitude of the magnetic field $B$ and the current bias $I_\mathrm{bias}$ in this few-mode regime, with low resistance supercurrent regions in dark blue around zero current bias. 

Note that the data at low field are asymmetric with respect to current reversal. Only one sweep direction is plotted for the rest of the figures.

A strong decrease of the switching current is observed from $B=\SI{0}{T}$ to $B=100-\SI{200}{mT}$. 
Beyond the initial decrease, the critical current exhibits nonmonotonic behavior with multiple nodes and lobes. 
Despite the \SI{1}{\micro \meter} contact separation, the supercurrent can be resolved up to fields as high as $B=\SI{2}{T}$, which is comparable to the estimated strength of the effective spin-orbit field $B_\mathrm{SO}$.
At finite magnetic fields where the Josephson energy is suppressed the sharp switching behavior is replaced with a smooth transition to a higher resistance state. 
In voltage-biased measurements, this manifests as a zero-bias conductance peak (see Supplemental Material ~\cite{supp}). 
This signal can mimic the onset of the topological phase since it is also associated with the zero-bias conductance peak that appears at a finite magnetic field.

\begin{figure}[t]
\includegraphics[width=\columnwidth]{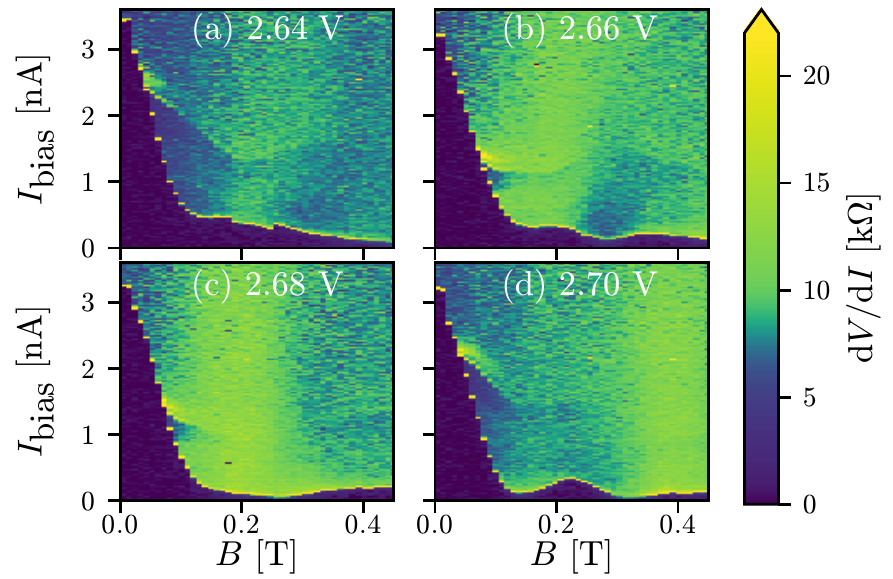}
\caption{(a)-(d) $\mathrm{d}V/\mathrm{d}I$ vs $B$ and $I_\mathrm{bias}$ for different gate voltage settings $V_\mathrm{g}$ indicated above each panel.  Data from device 2; see the Supplemental Material~\cite{supp} for the scanning electron micrograph of the device with the tuned gate marked.}
\label{fig:figure2}
\end{figure}

We now qualitatively discuss the possible explanations for the behavior observed in Fig. \ref{fig:figure1}(b).
Zeeman splitting can induce $0-\pi$-junction transitions which result in an oscillatory Josephson energy as a function of the magnetic field \cite{bulaevskii1977superconducting,  buzdin1982critical, demler1997superconducting}. 
This alternating $0-\pi$ junction behavior is due to spin-up and spin-down channels acquiring different phases as they travel across the junction [Fig. \ref{fig:figure1}(a)]. 
However, in our junctions a strong spin-orbit effective field, which has been reported to point perpendicular to the nanowire \cite{nadj2012spectroscopy}, reduces the relative phase shifts of spin-up and spin-down and lifts the nodes in the supercurrent\cite{shumeiko2008, yokoyama2014anomalous,yokoyama2014magnetic}.
For the spin-orbit strength previously reported in InSb nanowires \cite{nadj2012spectroscopy, vanweperen2015spinorbit}, we estimate an effective spin-orbit field $B_{\rm SO} \sim 1-\SI{2}{T}$ for a chemical potential value in the middle of the subband.
Therefore, we do not expect the occurrence of $0-\pi$-transitions in ballistic nanowires for fields much lower than this typical value of $B_{\rm SO}$, unless the chemical potential is close to a transverse mode edge (within $1-\SI{2}{meV}$), where $B_{\rm SO}$ is suppressed.
Given the typical mode spacing of $10-\SI{20}{meV}$~\cite{vanweperen2012quantized,kammhuber2016conductance}, in combination with the occurrence of several nodes well below \SI{1}{T}, the Zeeman $\pi$-junction effect is an unlikely explanation for all of the critical current nodes observed here for generic device settings.

Supercurrents carried by different transverse modes would also acquire different phase shifts and interfere due to mode mixing within the wire or at the contact between the nanowire and the superconductor lead~\cite{PhysRevB.91.245436}. 
Such interference is analogous to the Fraunhofer effect in wide uniform junctions: it becomes relevant when a single superconducting flux quantum is threaded through the nanowire cross section, a regime that is reached for $B \approx \SI{0.25}{T}$, well within the range of the present study. 
Comparison of the experimental and numerical data in this Letter suggests that this is the effect that dominates the magnetic field dependence of the critical current.

Transitions in and out of the topological superconducting phase in the nanowire segments covered by the superconductors were also predicted to induce reentrant critical current\cite{san2014mapping}. 
Although we used devices similar to those presented in recent Majorana experiments \cite{mourik2012signatures,Zhang2016ballisticmajoranadevice,chen2016phasediagram}, here we did not gate tune the regions of the wire underneath the superconducting contacts into the topological regime. An accidental topological regime occurring on both sides of the junction in multiple devices is an unlikely explanation for the generic observations reported here. 

Figure~\ref{fig:figure2} shows a typical sequence of magnetic field dependences of the critical current, obtained by adjusting one of the narrow local gates. 
The critical current exhibits multiple nodes [Fig.~\ref{fig:figure2}(d)], just a single node [Fig.~\ref{fig:figure2}(c)], or no node [Fig.~\ref{fig:figure2}(a)] in the same field range.
At some nodes the critical current goes to zero, while a nonzero supercurrent is observed at other nodes. 
No periodic patterns such as those characteristic of a dc-SQUID or a uniform junction are observed. 
Note that slight changes in the gate voltage are sufficient to dramatically alter the magnetic field evolution curve; the corresponding change in chemical potential $\Delta \mu$ is small ($\Delta \mu < \SI{1}{\milli \electronvolt}$) compared with the typical intermode spacing ($\sim \SI{15}{\milli \electronvolt}$). 
Furthermore, the gate used only tunes a \SI{100}{\nano \meter} segment of the \SI{650}{\nano \meter} long junction.

\begin{figure}
\includegraphics[width=\columnwidth]{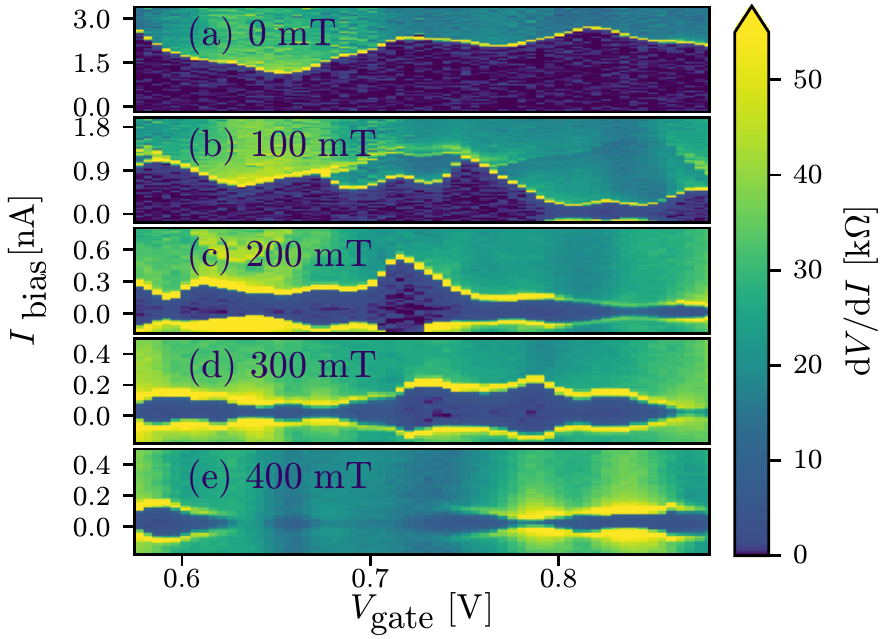}
\caption{(a)-(e) $\mathrm{d}V/\mathrm{d}I$ vs $V_\mathrm{g}$ and $I_\mathrm{bias}$ at different $B$ (indicated within each panel). Data from device 2. The gate used for tuning is different from that used in Fig.~\ref{fig:figure2}, see the Supplemental Material~\cite{supp}. }
\label{fig:figure3}
\end{figure}

Typical gate sweeps of the supercurrent are presented in Fig.~\ref{fig:figure3}. 
The critical current is strongly reduced at fields above \SI{100}{\milli \tesla} irrespective of the gate voltage. 
At all fields, the supercurrent is strongly modulated by the gate voltage. 
However, gate voltages at which nodes in the critical current occur differ for each magnetic field. 
Thus, no straightforward connection can be made between the zero-field critical current and node positions at a finite field, see also Fig.~\ref{fig:figure5}(a). 

\begin{figure*}[t]
\includegraphics[width=\textwidth]{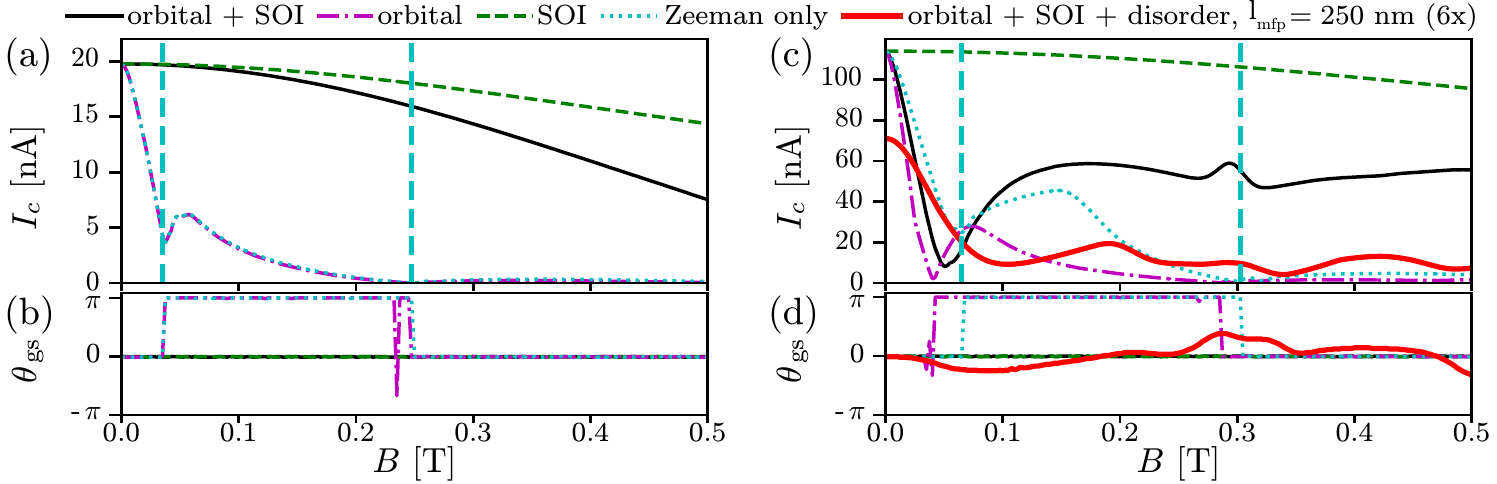}
\caption{Critical current and corresponding ground state phase difference for different combinations of terms in the Hamiltonian.
The Zeeman effect ($g=50$) is present in all of the curves.
Only the system corresponding to the red curve labeled with $l_\textrm{mfp}=\SI{250}{nm}$ includes disorder, for which the critical current is multiplied by a factor of 6. The simulation is performed at $T=\SI{100}{mK}$.
The curves in panels (a) and (b) are for a single spinful transverse mode ($\mu=\SI{10}{meV}$).
Panels (c) and (d) are for the multimode (three transverse or six spin-full modes) regime ($\mu=\SI{20}{meV}$).
The vertical thick dashed light blue lines in (a) and (c) indicate the positions of $0-\pi$ transitions in the absence of disorder and with $\alpha = 0$, $\bm{A}=0$.
Where not specified, the other constant simulation parameters are $\alpha=\SI{20}{\nm \meV}$, $m_\textrm{eff}=0.015 m_\textrm{e}$, $\Delta_\textrm{ind}=\SI{0.250}{meV}$; the lattice constant $a=\SI{8}{nm}$, the nanowire diameter $d_1=\SI{104}{nm}$, the outer diameter (with superconductor) $d_2=\SI{120}{nm}$, and the superconductor coverage angle (see the Supplemental Material~\cite{supp}, Fig.6) $\phi=135\si{\degree}$. For plots of corresponding current-phase relationships, Josephson energies, and numerical geometry, see the Supplemental Material ~\cite{supp}, Fig. 6-9.
}
\label{fig:critical_currents}
\end{figure*}

In order to understand the magnetic field evolution of the Josephson effect, we develop an effective low-energy model of a spin-orbit and Zeeman-coupled few-mode nanowire, covered by superconductors at both ends. 
We define $x$ as the direction along the wire, $y$ perpendicular to the wire in the plane of the substrate, and $z$ perpendicular to both wire and substrate. 
The corresponding Hamiltonian reads
\begin{align}
H = &\left(\frac{\bm{p}^2}{2m^*}-\mu + \delta U\right)\tau_z + \alpha (p_x \sigma_y - p_y \sigma_x)\tau_z \nonumber \\ &+ g \mu_B \bm{B}\cdot\boldsymbol{\sigma} + \Delta \tau_x. 
\label{eq:H}
\end{align}
Here $\bm{p}=-i\hbar\nabla+e\bm{A}\tau_z$  is the canonical momentum,  where $e$ is the electron charge, and $\bm{A}={\left[ B_y z - B_z y,\; 0,\; B_x y\right]}^{T}$ is the vector potential chosen such that it does not depend on $x$. 
Further, $m^*$ is the effective mass, $\mu$ is the chemical potential controlling the number of occupied subbands in the wire, $\alpha$ is the strength of Rashba spin-orbit interaction, $g$ is the Land{\'e} $g$-factor, $\mu_\mathrm{B}$ is the Bohr magneton, and $\Delta$ is the superconducting pairing potential.
The Pauli matrices $\sigma_i$ and $\tau_i$ act in spin and electron-hole spaces, respectively.
We assume that the electric field generated by the substrate points along the $z$ direction, such that the Rashba spin orbit acts in the $xy$\kern-.05ex-plane, which is at low energies equivalent to an effective magnetic field $\bm{B}_{\rm SO}\parallel\hat{y}$.
We include the vector potential in the tight-binding system using the Peierls substitution \cite{hofstadter_energy_1976}.
Finally, we include an uncorrelated on-site disorder $\delta U \in [-U, U]$, with $U$ the disorder strength, which we parametrize by a normal state mean free path $l_\textrm{mfp}$. \cite{beenakker1997random}$^\textrm{,}$\footnote{To determine $l_\textrm{mfp}$ we calculate a disorder-averaged normal state conductance $g$ and evaluate the mean free path $l_{mfp}$ by fitting, $g=g_0 N_\textrm{ch} / (1 + L / l_\textrm{mfp})$, with $N_\textrm{ch}$ the number of conduction channels, and $g_0$ conductance quantum.}

We perform numerical simulations of the Hamiltonian \eqref{eq:H} on a 3D lattice in a realistic nanowire Josephson junction geometry. 
The critical current is calculated using the algorithm described in Ref.~\onlinecite{ostroukh2016two} and the Kwant code \cite{groth_kwant}.
We note that for moderately damped and overdamped Josephson junctions, such as those studied here, the theoretical $I_c$ closely follows the experimentally measured switching current \cite{kautz1990noise}. 
The source code and the specific parameter values are available in the Supplemental Material ~\cite{supp}.
The full set of materials, including computed raw data and experimental data, is available in Ref.~\onlinecite{data}.

Numerical results are presented in Figs. \ref{fig:critical_currents} and \ref{fig:figure5}(b).
First, we discuss the case of only a single transverse mode occupied [Figs.~\ref{fig:critical_currents}(a) and ~\ref{fig:critical_currents}(b)], which is pedagogical but does not correspond to the experimental regime.
When all field-related terms of Eq.~\eqref{eq:H} are included ($\bm{A}\neq 0$, $\alpha \neq 0$), we observe a monotonic decay of the critical current much more gradual than in the experiment, due to the absence of the intermode interference effect in the single-mode regime.
The $\pi$-junction transitions do not appear up to fields of order $\SI{0.5}{T}$ due to the strong spin-orbit effective field, which keeps spin-up and spin-down at the same energy so that they acquire the same phase shifts traversing the junction.
The critical current eventually decays because the Zeeman term overtakes the spin-orbit term at fields greater than \SI{0.5}{\tesla}. 
When the spin-orbit term is turned off ($\alpha = 0$), we see several $0-\pi$ transitions taking place within the studied field range, confirmed by the ground state phase switching between 0 and $\pi$ at a series of magnetic fields [Fig.~\ref{fig:critical_currents}(b)].

The experimentally relevant regime is when several transverse modes are occupied. 
The measurements display three qualitative features: (i) the initial critical current at $B=\SI{0}{T}$ is strongly suppressed within $100-200$mT; (ii) the critical current then revives and continues to display nodes of variable depth and periodicity; (iii) this revival of the critical current after suppression is about 10\% of its original value at $B=\SI{0}{T}$.
Models that neglect the orbital effect display either a slow monotonic decay of the critical current (spin-orbit included, $\alpha \neq 0$), or regular critical current nodes due to $0-\pi$ transitions (no spin-orbit, $\alpha=0$) [Fig.~\ref{fig:critical_currents}(d)], as in the single-mode case.
When orbital effects are included, $\bm{A}\neq 0$, observations (i) and (ii) are reproduced but the revival of the critical current after initial suppression is still strong.
Inclusion of a realistic amount of disorder, which creates additional interference paths and suppresses supercurrent further, reproduces all observations (i), (ii), and (iii).
Thus, we conclude that the experiment is best reproduced when $\bm{A}\neq 0, \alpha \neq 0$ and weak disorder that induces intermode scattering is included within the junction model.

The inclusion of disorder in the multimode regime breaks mirror symmetry~\cite{yokoyama2014anomalous,yokoyama2014magnetic} and generates a spin-orbit field along the external magnetic field \textit{B}, which gives rise to a nonsymmetric current-phase relation, inducing a $\varphi_0$ junction (see the Supplemental Material ~\cite{supp}, Sec. VIII, for a detailed explanation). 
The ground state phase of the $\varphi_0$ junction can continuously change between 0 and $\pi$ [red trace in Fig.~\ref{fig:critical_currents}(d)]. 
Experimental verification of such phase-related effects is not possible in the two-terminal junction geometry used here, it requires phase-sensitive experiments in the SQUID geometry.

\begin{figure}[t]
\includegraphics[width=\columnwidth]{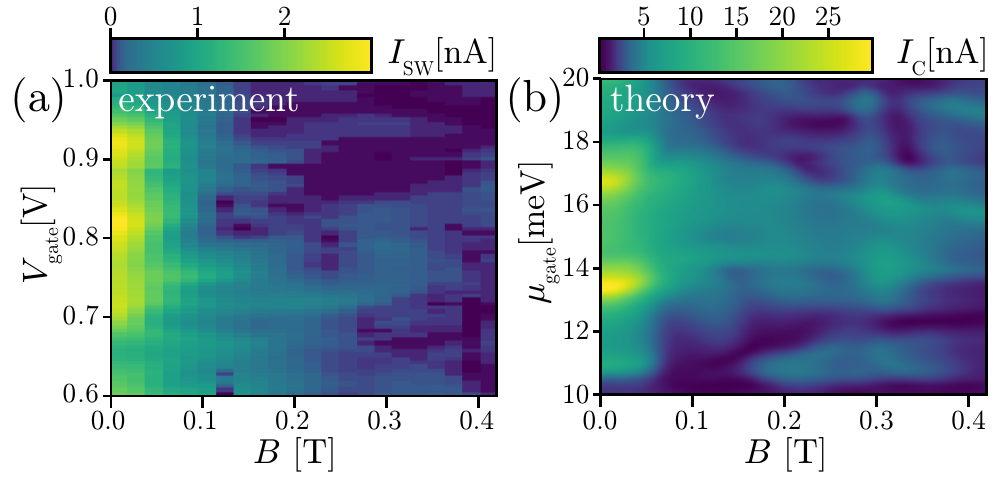}
\caption{Comparison between experimental (a) and numerical (b) results.
The parameters for the numerical simulations are the same as in Figs. \ref{fig:critical_currents}(c) and \ref{fig:critical_currents}(d), red curve.
The range of the chemical potential in the gated region ($\mu_{\textrm{gate}}$) is chosen using Ref. \onlinecite{vuik}. The experimental data are taken with device 2.}
\label{fig:figure5}
\end{figure}

In Fig.~\ref{fig:figure5} we compare side-by-side experiment and simulations via field-versus-gate maps of the supercurrent. 
In Fig.~\ref{fig:figure5}(a), the switching current from a set of $\mathrm{d}V/\mathrm{d}I$ vs. $I_\textrm{bias}$ traces similar to those in Fig.~\ref{fig:figure3} was extracted from device 2 (see the Supplemental Material ~\cite{supp} for algorithm details). 
Beyond the decay of the switching current on the scale of $\SI{100}{mT}$, the experimental data show a complex evolution of switching current maxima and minima in gate-field space. 
Characteristic features of this evolution are reproduced by our simulation shown in Fig.~\ref{fig:figure5}(b).
In particular, the experimentally observed magnetic field scale of initial supercurrent decay is reproduced in the simulation.
Furthermore, the gate-tunable maxima and minima of the critical current are recovered in our model; both in experiment and simulation these do not evolve in a regular fashion (a consequence of the complexly shaped interference trajectories). 
This qualitative agreement found additionally substantiates the applicability of our model to the experimental results.

Our results are instrumental for modeling Majorana setups. 
Specifically, the decrease of Josephson energy by an order of magnitude is observed at fields at which the onset of topological superconductivity is reported. 
This effect should, therefore, be taken into account in efforts to realize recent proposals for fusion and braiding of Majorana fermions \cite{Hyart2013braiding, Aasen2016briading, plugge2016majorana, Karzig2016braiding}, especially in those that rely on controlling the Josephson coupling \cite{Hyart2013braiding, Aasen2016briading, plugge2016majorana}. 
Our findings are applicable not only to bottom-up grown nanowires and networks but also to scalable few-mode junctions fabricated out of two-dimensional electron gases. \cite{nichelle2017scalingzbp,lee2017inas2deg} 
We suggest that in such devices narrow multimode nanowires should be used. 
At the magnetic field strengths required for braiding the many modes would facilitate strong Josephson coupling, whereas a small diameter prevents its suppression due to supercurrent interference.

Phase-sensitive measurements in the SQUID loop geometry will reveal effects  such as the Zeeman-induced  $\pi$ junction and the spin-orbit induced $\varphi_0$ junction, which our study identifies numerically but does not access experimentally. 
Single quantum mode junctions are within reach thanks to the recent demonstration of quantum point contacts in InSb nanowires at a zero magnetic field \cite{kammhuber2016conductance}. 
In that regime phenomena such as induced $p$-wave superconductivity can be studied in a unique gate-tunable setup, when tuning down to a single spin-polarized mode in the weak link. 
The results are also applicable to other interesting material systems where spin-orbit, orbital,  and Zeeman effects interplay - systems such as Ge/Si, PbS, InAs, and  Bi nanowires and carbon nanotubes. \cite{cleuziou2006carbon}

\textit{Acknowledgments.} This work has been supported by the European Research Council, the Netherlands Organization for Scientific Research (NWO), the Foundation for Fundamental Research on Matter (FOM), and Microsoft Corporation Station Q.
V.M. was supported by a Niels Stensen Fellowship for part of the research. 
A.R.A., D.I.P., and S.M.F. are grateful to KITP, where part of the research was conducted with the support of NSF Grant No. PHY11-25915. S.M.F acknowledges NSF Grant No. DMR-125296, S.M.F., L.P.K. and E.P.A.M.B. acknowledge ONR. 
D.I.P. thanks NSERC, CIFAR, and the Max Planck - UBC Centre for Quantum Materials for support.

\onecolumngrid
\clearpage
\begin{center}
\textbf{\large Supplemental Material: Supercurrent interference in few-mode nanowire Josephson junctions}
\end{center}

\setcounter{equation}{0}
\setcounter{figure}{0}
\setcounter{table}{0}
\setcounter{page}{1}
\makeatletter
\renewcommand{\theequation}{S\arabic{equation}}
\renewcommand{\thefigure}{S\arabic{figure}}

\section{Zero field gate dependence}

\begin{figure}[!h]
\centering
\includegraphics[width=\textwidth]{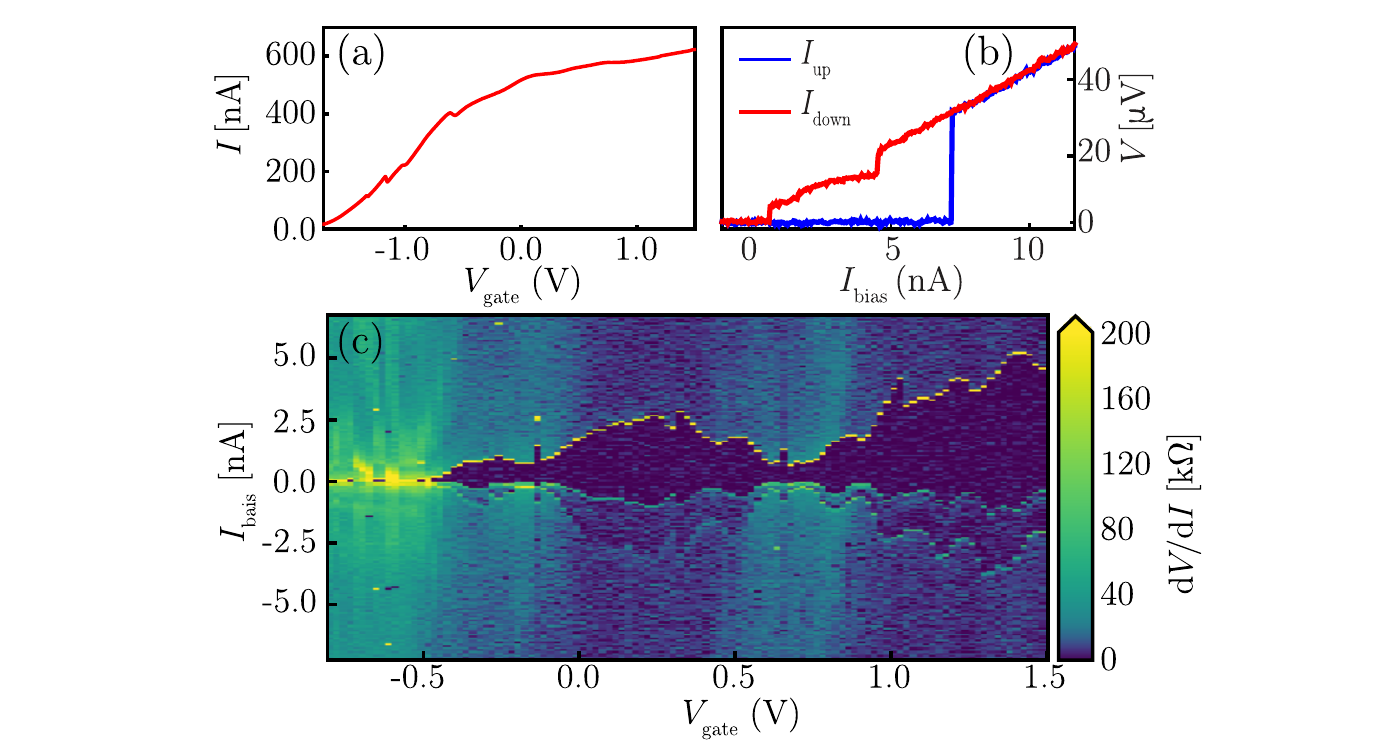}
\caption{(a) Device current as a function of gate voltage, $V_\mathrm{bias}$ = 10 mV. Except the gate that is varied in this scan, other gates are at +3 V. (b) Voltage-current characteristic for both upwards (blue) and downwards (red) sweeping direction of current bias. The supercurrent of 8 nA is the maximum supercurrent observed in this device and corresponds to all gates at +3V. (c) Numerical derivative $\mathrm{d}V/\mathrm{d}I$ of $V\left(I\right)$ as function of current and gate voltage. Current bias is swept from negative to positive.}
\label{fig: charaterization_D1}
\end{figure}

Characterization of device 2 at $B$=0 T is shown in Figure \ref{fig: charaterization_D1}, devices 1 and 3 behave similarly (data not shown).
Current versus local gate voltage is measured at $V_{\text{bias}}$ = 10 mV (Figure \ref{fig: charaterization_D1}(a)). 
Taking known series resistances into account, the device resistance of $\sim$6 k$\Omega$ is found, corresponding to the sum of the conduction channels and contact resistances.

As shown in Figure \ref{fig: charaterization_D1}(b), by optimizing the gate voltages a maximal supercurrent of 8 nA was found, with a corresponding voltage of 32 $\mu$V, which developed upon switching to the normal state. 
The junction is hysteretic as shown by the low retrapping current, and has a sharp transition to the normal state, indicating that the junction is in the underdamped regime. 
Note that self-heating may also contribute to the hysteresis \cite{pekola2008hysteresis}.

\section{Shapiro step measurements}

\begin{figure}[!h]
\centering
\includegraphics[width=\textwidth]{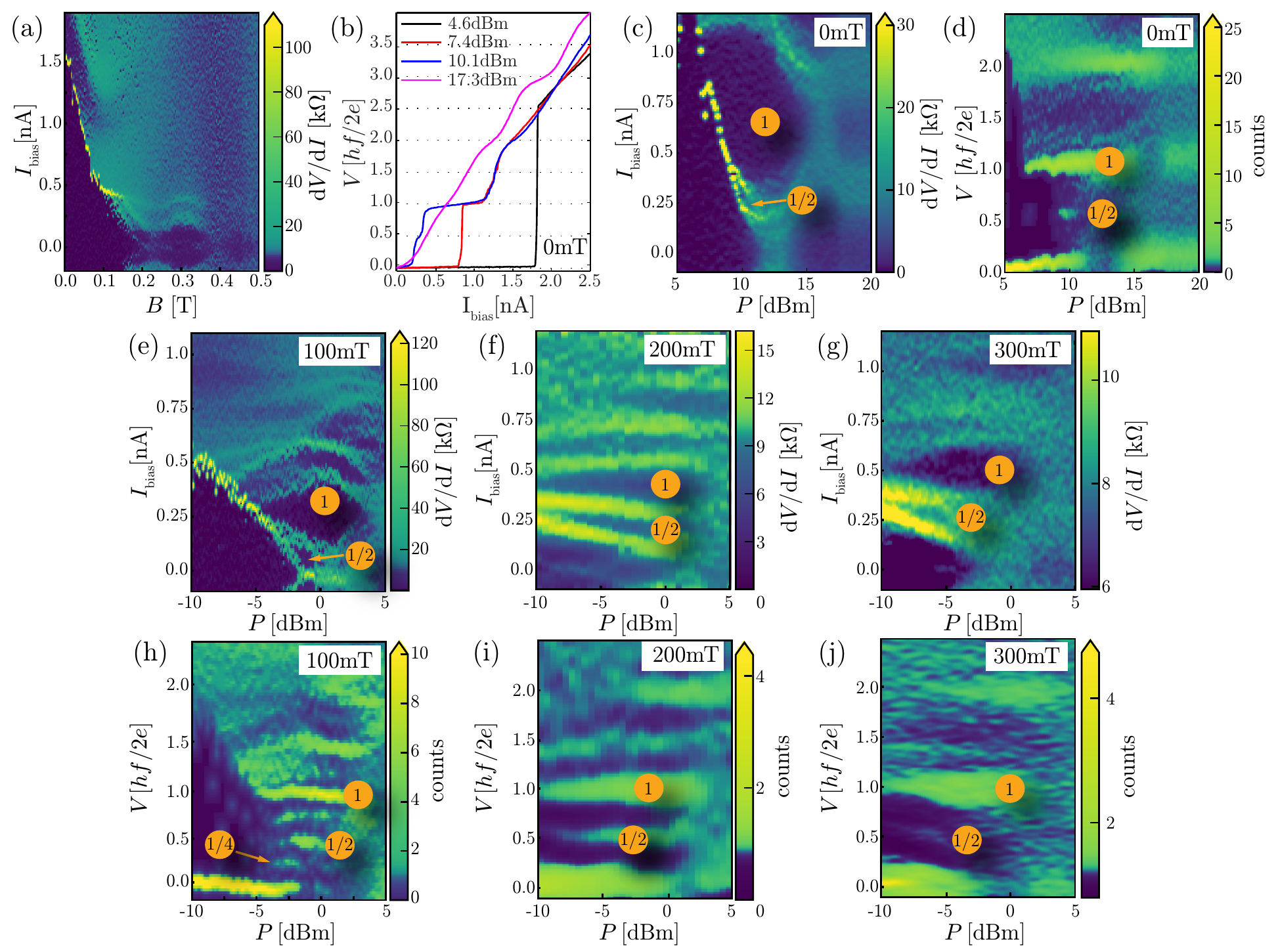}
\caption{
Shapiro steps in magnetic field. 
(a) $B$ dependence of supercurrent without microwave radiation applied. 
Numerical derivative $\mathrm{d}V/\mathrm{d}{I}$ of the original $V\left(I\right)$ curves is shown. 
(b) Shapiro steps at $B = \SI{0}{\tesla}$ for different microwave powers. 
At the lowest RF power of 4.6 dBm (black line) no Shapiro steps are present. 
A half integer step is visible at 10.1 dBm (blue line). 
(c),(e)-(g) Microwave power dependence of Shapiro steps for different $B$. Numerical derivative $\mathrm{d}V/\mathrm{d}{I}$ of the original $V\left(I\right)$ curves is shown, in this representation the Shapiro step plateau corresponds to low differential resistance (blue color).
(d), (h)-(j) are Shapiro steps for different $B$ as a function of microwave power plotted in histogram. High voltage counts correspond to the plateaus in the Shapiro steps.
At $B = \SI{0}{\tesla}$ [panel (c)], the power dependence is dominated by integer Shapiro steps and only a small contribution of half integer steps is visible. 
At $B = \SI{100}{m\tesla}$ [panel (e)]  fractional steps are visible. Noticeably, not only half integer steps, but also quarter steps are weakly present in panel (h). 
$B = \SI{200}{m\tesla}$ [panel (f)] is closest to the minimum supercurrent at \SI{250}{m\tesla}. Here the half integer and integer steps are almost equal in width. 
Finally, beyond the minimum of supercurrent, at $B = \SI{300}{m\tesla}$ [panel (g)], the integer steps increase again in width relative to the half integer step.
Curves in (b) are from the same dataset as shown in (c). 
Values given for the RF power in panels (b)-(j) are the output power of microwave source, \SI{60}{\deci \bel} attenuation, of which \SI{20}{\deci \bel} at low T, is applied. 
Data is from device 2, second cooldown.}
\label{fig: shapiro}
\end{figure}

Device 2 has been cooled down a second time with a microwave antenna near the sample. 
This enabled the study of Shapiro steps in the junction as a function of microwave power and frequency, see Fig.~\ref{fig: shapiro}.
The device is again tuned to a multi-mode regime, comparable to $V_\mathrm{gate}$ = 0.5 V in Fig.~\ref{fig: charaterization_D1}(c). 
Due to an increased microwave background noise in the vicinity of the junction upon adding the antenna, an extra rounding of the $V\left(I\right)$-trace near the switching bias is present. 

Figure \ref{fig: shapiro}(a) is a magnetic field $B$ dependence of supercurrent in the absence of microwave drive. 
The supercurrent pattern as a function of $B$ is similar to the one shown in Figure 2 of the main text. 
This indicates that thermally cycling the device did not change the qualitative behavior of the device, although the exact gate tunings are different. 

We focus on the power dependence of Shapiro steps at different $B$ strengths of 0mT, 100mT, 200mT and 300 mT corresponding to Fig.~\ref{fig: shapiro}(c), (e), (f), (g) respectively. 
The microwave frequency is kept fixed at 2.0 GHz. Shapiro steps show up at voltages corresponding to $V=n\cdot\frac{hf}{2e}$, where $n$ may be a fraction. At $B$ = 0 mT [Fig.~\ref{fig: shapiro}(c)], half integer steps are only weakly present. 
At $B$ = 100 mT [Fig.~\ref{fig: shapiro}(e)], not only $n$ = 1/2 steps but also weak $n$ = 1/4 steps are visible (not marked with circled number). 
This is clearly visible in Fig.~\ref{fig: shapiro}(h), where the same data is plotted in a voltage histogram, with high voltage counts corresponding to the plateaus of the Shapiro steps.

The $B$ = 200 mT and $B$ = 300 mT cases [Fig.~\ref{fig: shapiro}(f), (g)] correspond to low critical currents. Nevertheless, Shapiro steps can still be resolved. At B = 200 mT, which is closest to the minimum of critical current, the width of the 1/2 step is more than half the width of the 1st step, and it is similarly large compared to the 1st step at 300 mT. Fig.~\ref{fig: shapiro}(i), (j) are the histogram representations of Fig.~\ref{fig: shapiro}(f), (g).

Shapiro steps at fractional frequencies, especially the half-integer steps, have been previously observed in Josephson junctions under various conditions\cite{Lehnert1999fractional,Dubos2001fractional, dinsmore2008fractional,PhysRevLett.92.257005,PhysRevLett.97.067006}. 
For instance, they can arise due to Josephson coupling of higher orders accompanied by a non-sinusoidal current-phase relationship\cite{PhysRevB.63.214512}. 
In quasi-ballistic few-mode Josephson junctions the current-phase relation is expected to be non-sinusoidal, consistent with half-integer Shapiro steps observed here even at zero magnetic field. 
The higher order 1/4-steps are more exotic and deserve a deeper study in the future, though they may also originate from a non-sinusoidal current-phase relationship. 
Non-sinusoidal current-phase relationships are obtained within our model, see Fig. \ref{fig:CPR}. However, Fourier analysis of the simulation suggests that Shapiro steps at 1/3 the Josephson voltage should dominate over 1/4 steps. This discrepancy remains not understood.

In a non-sinusoidal Josephson junction tuned to the $0-\pi$ transition the first order Josephson effect which is responsible for strong integer Shapiro steps vanishes, thus the current phase relationship is dominated by higher harmonics. 
In this case, Shapiro steps at half-integer and integer frequencies are expected to appear with the same step widths. 
The results presented here show that the ratio of step widths for half integer to integer steps indeed increases near a field-induced node in the critical current. However, the results are not conclusive as to whether this is due to a $0-\pi$ transition.

On the other hand, Majorana zero modes coupled across a junction barrier are predicted to result in disappearing odd-integer Shapiro steps\cite{lutchyn2010majorana,oreg2010helical}. 
Thus the behavior observed here is opposite to that expected due to Majorana modes: extra fractional steps in addition to integer steps are observed.

\newpage
\section{Angle dependence of fluctuations}
\onecolumngrid
In this section we present results from device 3 [Fig.~\ref{fig: SEM_D1D3}(c)] with contact spacing of 150 nm on which we performed current bias measurements with similar conditions as reported in the main text. 
Device 3 is fabricated with similar methods as Device 1 and 2, with the exception that $\text{HfO}_x$ is used as the dielectric material instead of $\text{SiN}_x$.

\begin{figure}[h]
\centering
\includegraphics[width=\textwidth]{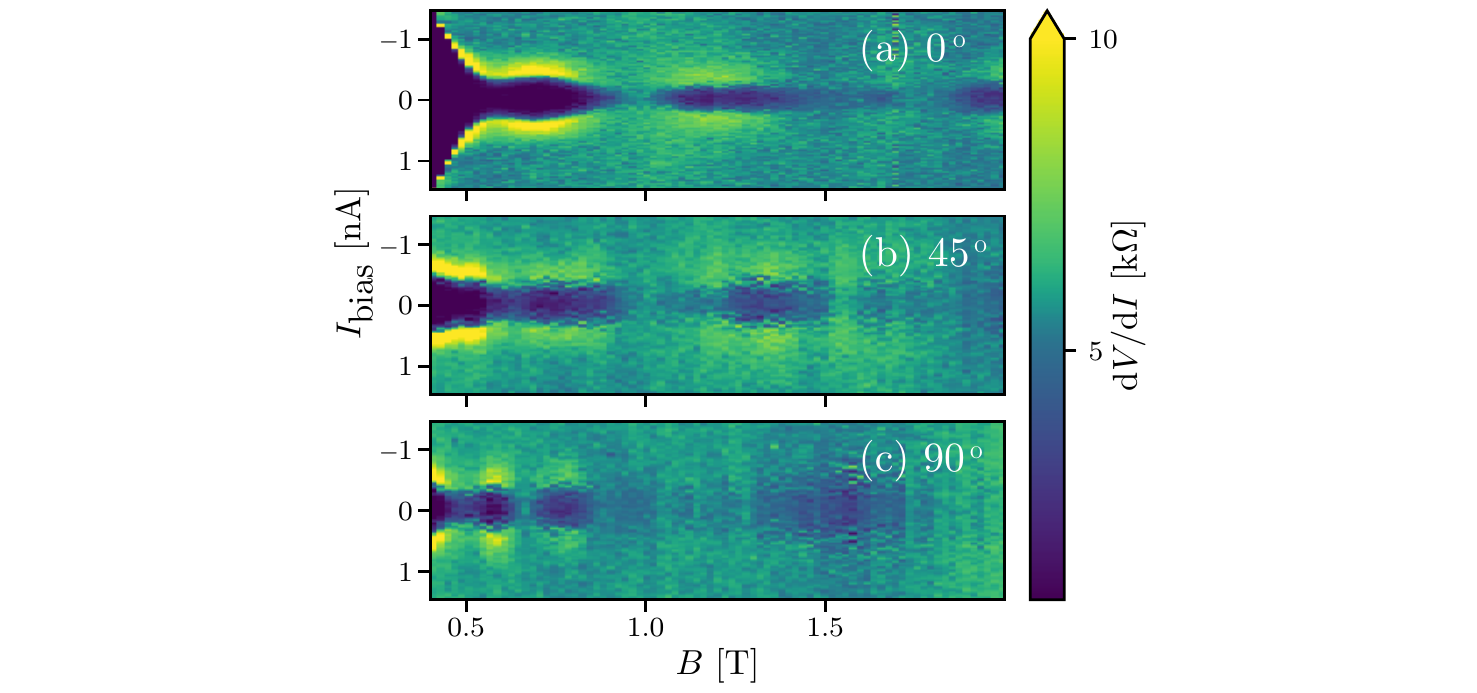}
\caption{Differential resistance measured as a function of current bias and magnetic field strength of Device 3. The angle indicated in each panel is the angle of the magnetic field relative to the wire axis, in the plane of the substrate.}
\label{fig:angle_dependence}
\end{figure}

The device shows a monotonic decrease of the critical current for magnetic field values up to 400 mT (not shown in the figure). This extended initial decay is attributed to the shorter contact separation, and hence reduced influence of disorder on intermode interference. 

Beyond 400 mT, the critical current fluctuates at a period depending on the direction of the magnetic field. 
Figure \ref{fig:angle_dependence} shows the differential resistance of the device for three different field directions. 
The top panel shows data where the field is pointed along the nanowire. 
The critical current decays until the field reaches 600 mT, beyond which it exhibits a weakly pronounced maximum and disappears at 900 mT after which it reappears again. 
As the field angle is rotated in the plane of the substrate [Figs.~\ref{fig:angle_dependence}(b),(c)], the critical current decays faster as a function of the field strength, and the subsequent nodes of the critical current are closer spaced in field. 
We associate this behavior with increased flux through the nanowire at finite angles between the field and the wire.

\begin{figure}[!h]
\centering
\includegraphics[width=\textwidth]{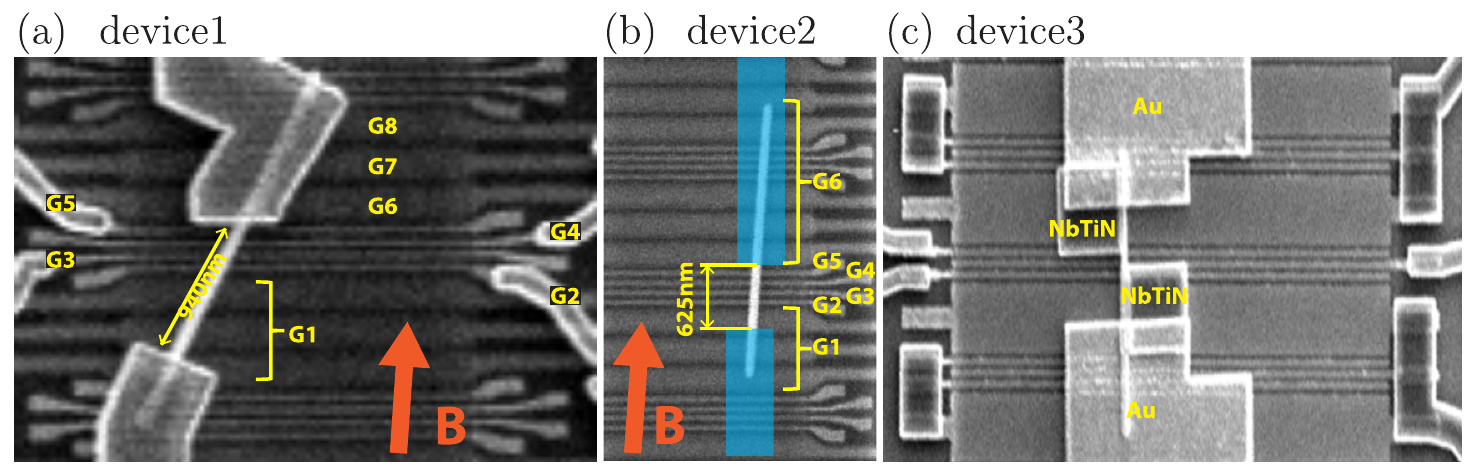}
\caption{Schematics based on SEM picture of device 1, device 2 and device 3. 
(a) Device 1, with an angle of $25^{\circ} \pm 5^{\circ}$ with the magnetic field. 
In all devices, not all local gates are operated independently: as indicated in the figures, larger gates are formed by shorting some of the local gates together, e.g G1. 
(b) Device 2, shown with the superconducting electrode design superimposed on top of the SEM image, as this device has not been imaged after the final fabrication step. 
The device has a contact spacing of $\sim \SI{625}{nm}$, with the wire at an angle of $0^\circ \pm 5^\circ$ with respect to magnetic field. 
(c) Device 3, incorporating two quasi-particles traps (Au) next to the superconducting contacts. 
The length of the Josephson junction is $\sim \SI{150}{\nano \meter}$. 
Device 3 is cooled down in a setup where the magnetic field could be rotated using a 3D vector magnet.}
\label{fig: SEM_D1D3}
\end{figure}

\newpage

\section{Zero bias peaks due to supercurrent can onset at finite magnetic field}

If the Josephson junctions are tuned into the topological regime, devices used in this study can also support Majorana fermions. 
As a matter of fact, such a design is employed by several groups for the purpose of searching for Majorana zero modes. 
Here we show that such Josephson junction based devices, even if the contacts are almost \SI{1}{\micro \meter} apart, cannot be used for unambiguous detection of Majorana zero modes \cite{deng2012ZBP, deng2014parity, Harlingen2013ZBP}. 
Specifically, we observe that, in a voltage-biased measurement, supercurrent can appear as a zero-bias peak that onsets at a finite magnetic field, in the same range of parameters as those used in Majorana experiments, thus mimicking a key Majorana signature.

\begin{figure}[!h]
\centering
\includegraphics[width=\textwidth]{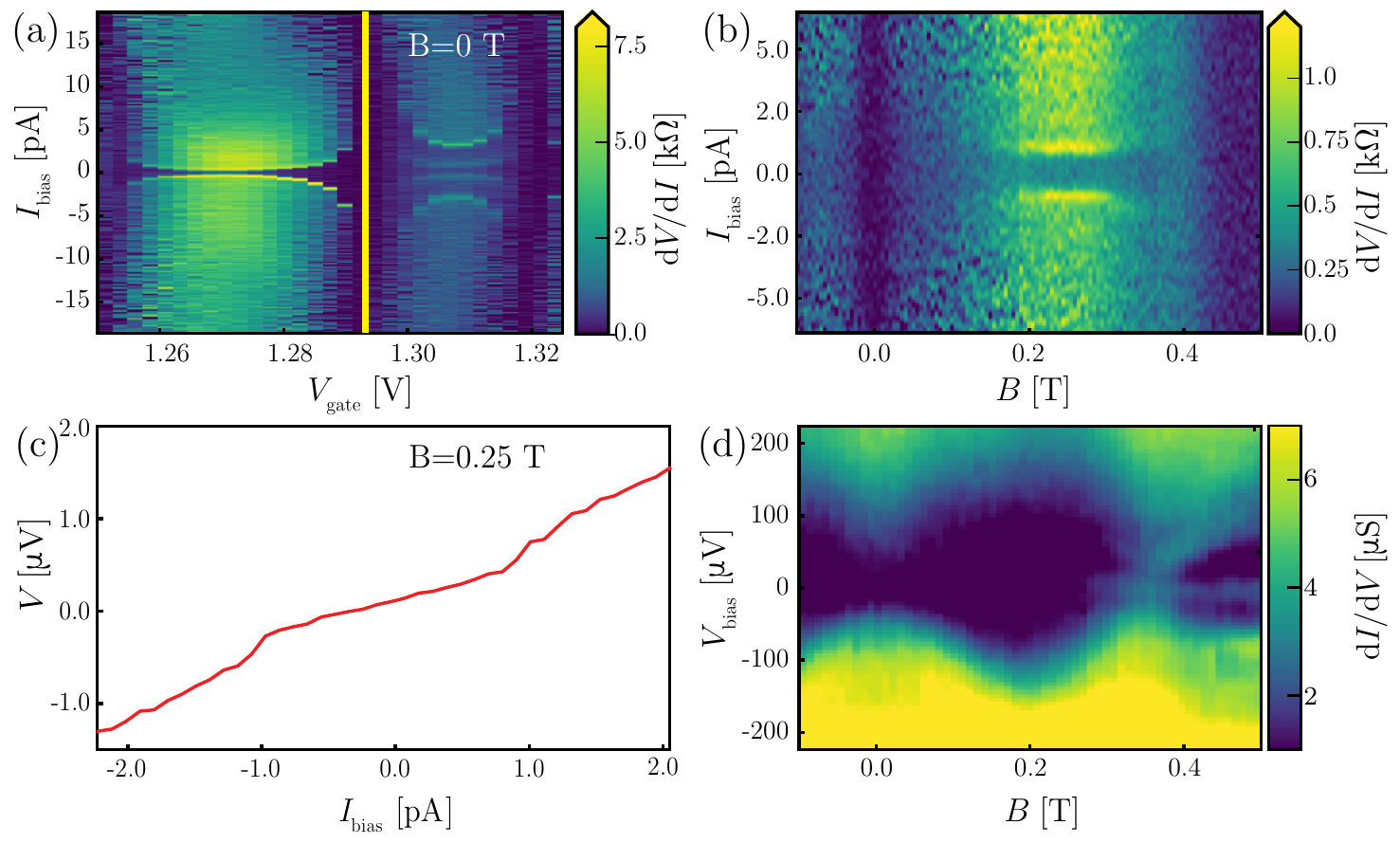}
\caption{Supercurrents and zero bias peaks at finite $B$.
(a) Differential resistance vs gate. In this scan, one of the local gates is set at -0.45 V and all other gates are at +1.5 V.
(b) Differential resistance vs $B$ at the indicated gate position in (a).
(c) Linecut from (b) at $B = \SI{0.25}{T}$. (d) Differential conductance vs $B$ corresponding to (b). Numerical derivative of original $V\left(I\right)$ curves is shown in (a) and (b). Data from device 1.}
\label{fig: supercurrentZBP}
\end{figure}

Figure \ref{fig: supercurrentZBP} shows the results. 
By applying a negative voltage to one of the local gates in between the superconducting contacts, a tunneling regime comparable to $V_\mathrm{gate}$ < -0.5 V shown in Figure \ref{fig: charaterization_D1}(a) for device 2 is achieved. 
The result of a current biased measurement in this regime is shown in Figure 5(a), a very small (down to 1 pA) supercurrent could be resolved. 
Interestingly, for gate regimes with lower resistance the supercurrent initially grows as expected, but then the $\mathrm{d}V/\mathrm{d}I$ peak related to the switching current broadens and is no longer visible. 
Here, we focus on the $B$ dependent behavior as shown in Figure \ref{fig: supercurrentZBP}(b),(c),(d) at a gate voltage indicated by the yellow line in Figure \ref{fig: supercurrentZBP}a. 
At $B=\SI{0}{T}$, no supercurrent was resolved in a current biased measurement, but upon increase of magnetic field, at around 200 mT, a small supercurrent shows up in a slightly more resistive regime. 
Such a small supercurrent may show up in a differential conductance measurement as a small zero bias peak (ZBP). 
Indeed, upon switching to a voltage biased differential conductance measurement, a small ZBP with height $\sim 0.01\frac{2e^2}{h}$ is found. 
Note that the ranges in which the supercurrent is visible in a current biased measurement and in which the ZBP is visible in a voltage biased measurement are not identical due to a minor charge switch between the two measurements.

\section{Extracting switching current from experimental data}

To obtain Fig. 5(a) of the main text, switching currents are extracted from a large set of voltage-current characteristics by numerically detecting the voltage step upon switching from the superconducting to the resistive regime. 
First, an initial low-pass filter is applied to the data reducing spurious fast fluctuations.
Next, a numerical derivative of the $V\left(I\right)$-curve is taken. 
This first derivative has a clear maximum for an $V\left(I\right)$-curve with a sharp transition, allowing for straightforward identification of the switching current.
However, the finite $B$-field $V\left(I\right)$-curves typically display smooth transitions from the superconducting to the resistive state, resulting in unclear or even absent maximums in the first derivative. 
A smooth transition still generates a maximum of the second derivative, allowing for identification of the switching current. 
We, therefore, introduced a threshold for a first derivative maximum, below which a second derivative is taken of the $V\left(I\right)$-curve with its maximum identified as the switching current.
A second threshold is introduced for the maximum of the second derivative, below which the switching current is considered to be zero. 
Algorithm parameters are optimized to both correctly identify the sharp transitions of large switching currents and to avoid false positives of small switching current.

\section{Details of the modeling}

We discretize the Hamiltonian (1) of the main text on a cubic lattice with a lattice constant of $a=\SI{8}{nm}$.
The nanowire cross section has a diameter of $\SI{104}{nm}$ and the superconductor on top of the semiconductor nanowire adds two more layers of unit cells partially covering the nanowire ($135^{\circ}$ of the wire's circumference). There are 3 free parameters in the simulation for obtaining the correct induced gap in the nanowires, namely the coverage angle of the superconductor, the tunnel barrier between the SC and the SM, and the superconducting gap. The coverage angle is fixed at $135^{\circ}$ in order to save computational time. Since the Meissner effect is not included in the simulation, the exact value of the angle does not play a critical role. The superconducting order parameter $\Delta$ is set such that the induced gap inside the nanowire at zero field is $\Delta_\textrm{ind} = \SI{0.250}{meV}$.

The superconductor has the same lattice constant and effective mass as the nanowire, justified by the long-junction limit.
This means that the wave function has most of its weight in the nanowire and that the superconducting shell merely serves as an effective boundary condition that ensures that all particles are Andreev-reflected.
Further, the superconductor lacks the Zeeman effect and spin-orbit interaction. Zeeman effect in the superconductor is neglected because the g-factor in NbTiN is 2, much smaller than the g-factor in InSb (which is 50). 
We use realistic parameters of an InSb nanowire~\cite{mourik2012signatures}:  $\alpha=\SI{20}{\meV\cdot\nm}$, $m^{*}=0.015 m_e$, and $g=50$. 

\begin{figure}[!h]
\centering
\includegraphics[width=\textwidth]{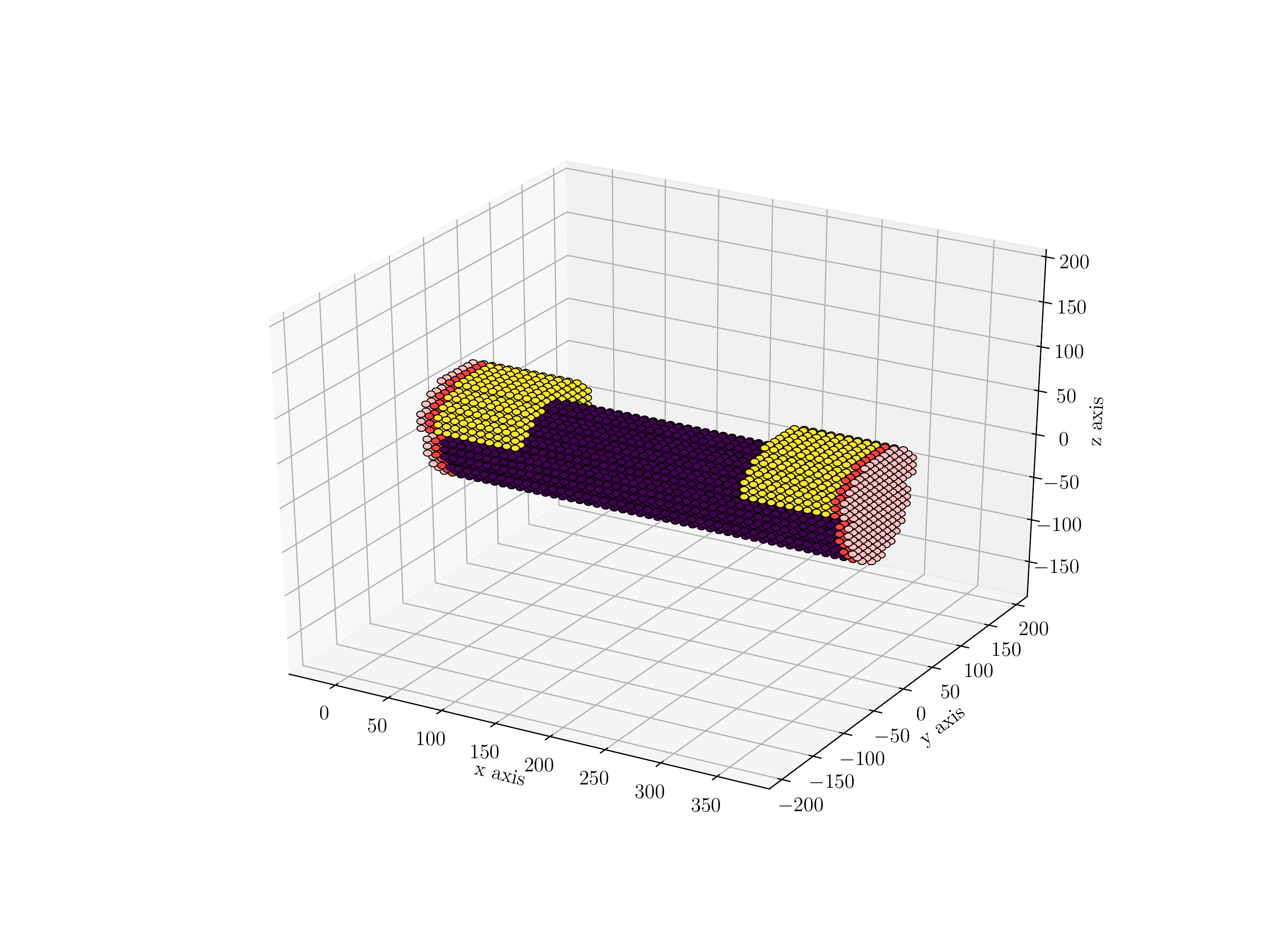}
\caption{The modeled tight-binding system.
The purple sites indicate the semi-conductor and the yellow sites show the superconductor.
The red and light red colored cross sections indicate that the wire extends infinitely in that direction.
We defined the length of the wire $L$ as the part that is not covered with the superconductor. In this figure for clarity we plot a shorter wire ($L=\SI{200}{\nm}$), while in the simulations we chose $L=\SI{640}{\nm}$.
The other dimensions used in the simulations are as depicted. 
Specifically, the wire diameter is $\SI{104}{\nm}$, the thickness of the superconductor is $16-\SI{24}{\nm}$, and the coverage angle of the superconductor is 135$^{\circ}$.}
\label{fig: system}
\end{figure}

The geometry of the modeled system is shown in Fig.~\ref{fig: system}.

\section{Detailed theoretical estimates}
In this section we estimate the strength of different possible mechanisms that can cause supercurrent fluctuations in the nanowire Josephson junction.

\textit{Interference between orbital channels}.
The area of the cross section of the nanowire is $\sim \pi \times (\SI{50}{nm})^2$. This means that the magnetic field value of $B \approx \SI{0.26}{T}$ corresponds to one flux quantum penetrating the cross section of the nanowire.
At this value of the magnetic field we expect the phase shifts between different bands propagating between the two superconductors to be comparable to $\pi$.
This sets the typical $B$ scale for the interference of different orbital modes carrying current, which is well within the experimentally observed typical difference in $B$ of consecutive critical current minimums.
This simple estimate neglects the magnetic field expulsion of the superconductor, which may create a higher flux in the nanowire near the superconducting contacts, thus lowering the effective field scale.

These estimates are similar to the analysis for the Fraunhofer-like interference in diffusive many-channel junctions\cite{Cuevas2007}. The novelty is, however, in the small number of channels in our junction, which causes irregular interference instead of the regular Fraunhofer pattern in the former case. Another important observation in our case is that even though the magnetic field is along the junction, it can still cause the interference due to different transverse profiles of the propagating modes.\\

\begin{figure}
\centering
\includegraphics[width=0.75\textwidth]{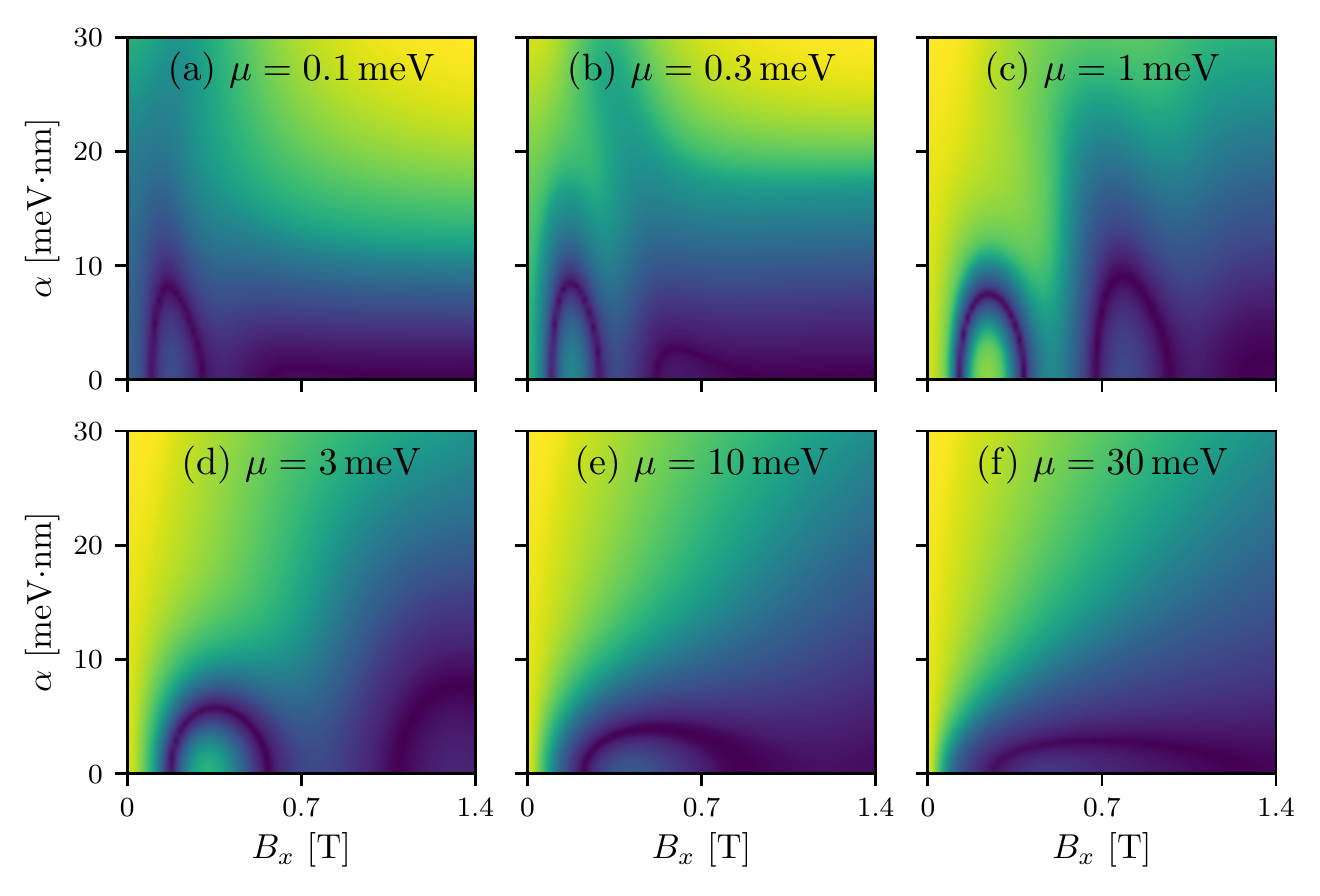}
\caption{Critical currents in a simple one-channel toy nanowire model\cite{lutchyn2010majorana,oreg2010helical} as a function of spin-orbit coupling strength $\alpha$ and magnetic field along the wire $B_x$.
The different panels (a)-(f) are taken at different chemical potentials, $0.1,\; 0.3,\; 1,\; 3,\; 10,\; \SI{30}{meV}$ respectively.
The color scales are not normalized across the different panels, but are all separately scaled to optimally show all features in every plot.
We observe how the $0-\pi$ transitions at finite $B_x$ get mutually annihilated upon increasing $\alpha$.\label{fig:currents_1D_alpha_vs_B_x}}
\end{figure}

\textit{Interference between spin channels}.
Supercurrent fluctuations can be produced by $0-\pi$ transitions due to the Zeeman splitting of the Andreev bound states inside the Josephson junction.
The characteristic $B$ scale of such supercurrent fluctuations is determined by the ratio of Zeeman energy to the Thouless energy.
This sets the relative phase $\theta_\mathrm{B}$ of the Andreev bound states, $\theta_\mathrm{B} = E_\mathrm{Z} L / \hbar v_\mathrm{F}$.
Here $E_\mathrm{Z}$ is the Zeeman energy, $L$ the length of the nanowire junction, and $v_\mathrm{F}$ the Fermi velocity in the nanowire.
The junction undergoes a $0-\pi$ transition when the relative phase difference of the ABS $\theta_\mathrm{B}$ reaches the value $\pi/2$. Such a transition is marked by a minimum in the junction critical current as a function of $B$.
Since $v_\mathrm{F}\approx \sqrt{2\mu / m^*}$, the field value at which $\theta_\mathrm{B}=\pi/2$ depends on the chemical potential $\mu$.
We thus estimate the upper bound of the magnetic field at which the first $0-\pi$ transition occurs by assuming a maximal value of $\mu \sim \SI{15}{meV}$ corresponding to the intermode spacing\cite{vanweperen2015spinorbit,kammhuber2016conductance}.
Assuming a junction length of $L=\SI{1}{\micro\meter}$, the upper bound of the transition occurs at $B\sim \SI{0.5}{T}$.
Generally, for smaller $\mu$, this value is significantly lower, therefore purely Zeeman induced supercurrent fluctuations are well within the range of our experiment.
These estimates are confirmed in our numerical simulations, see $\alpha=0$ lines of Fig.~\ref{fig:currents_1D_alpha_vs_B_x}.

\textit{Interference between spin, Zeeman and spin-orbit}.

The previous discussion on spin related interference considered the Zeeman effect only.
However, the strong spin-orbit interaction in the nanowire fixes the spin direction to the propagation direction and thus counteracts the effect of the Zeeman splitting. 
Following Ref.~\onlinecite{yokoyama2014anomalous}, the characteristic parameter for spin-orbit is $\theta_{SO} = \frac{\alpha k_\mathrm{F} L}{\hbar v_\mathrm{F}} = \frac{\alpha m^* L}{\hbar^2} = L/L_\mathrm{SO}$.
Here $L_\mathrm{SO}$ is the spin-orbit length, which is expected to be in the $50-\SI{250}{nm}$ range, much shorter than $L$.
For the Zeeman effect to cause a $0-\pi$ transition it needs to overcome the spin-orbit spin quantization. 
This means that the spin-orbit term increases the field at which the first $0-\pi$ transition happens, and this increase is stronger as the chemical potential is further away from the band bottom.
This interplay between Zeeman and spin-orbit interaction is expected to be highly anisotropic\cite{yokoyama2014anomalous} in the direction of $B$; the scenario described above assumes the external $B$ field and effective spin-orbit field to be perpendicular, as is expected for applying $B$ along the nanowire axis.
To substantiate our estimates we have used a nanowire toy model\cite{lutchyn2010majorana,oreg2010helical} to obtain critical current as a function of gate voltage, magnetic field, and spin-orbit coupling in Fig.~\ref{fig:currents_1D_alpha_vs_B_x}.
The model indeed illustrates that the further the chemical potential is from the bottom of the band the higher is the value of the magnetic field at which the $0-\pi$ transition occurs. 

In summary, the above estimates suggest that orbital interference is present regardless of the exact value of $\mu$, whereas spin related interference is highly restricted in $\mu$ range.
This favors an orbital interference interpretation of the experimental observations, since the supercurrent variations in the experiment are always present in a similar field range no matter the exact gate potential.

To illustrate this reasoning we produced Fig.~\ref{fig:currents_1D_alpha_vs_B_x}, which shows supercurrent fluctuations as a function of the distance to the bottom of the band in a single-band wire. With increasing the distance to the bottom of the bands $0-\pi$ transitions happen at higher fields. Upon ramping up spin-orbit strength the $0-\pi$ transitions disappear.

\section{Current phase relations and Josephson energies}
\begin{figure}[!h]
\centering
\includegraphics[width=\textwidth]{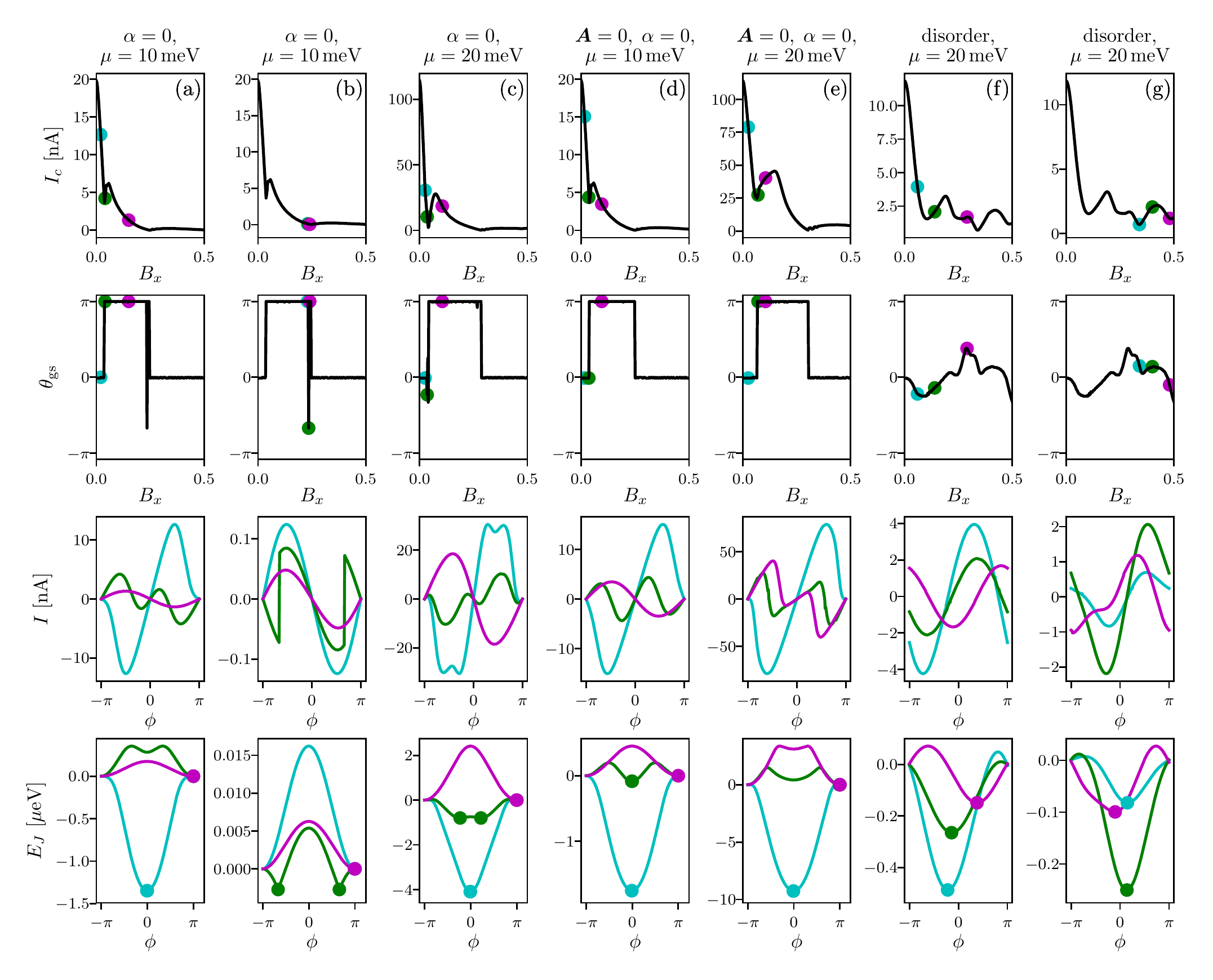}
\caption{(a)-(g) The critical current and ground state phase difference as function of magnetic field, and current phase relations (CPRs) and Josephson energies as functions of phase difference between the superconductors, for different parameters used in the model as labeled.
From top to bottom: $I_c(B_x)$ and $\theta_{\textrm{gs}}$ with the three points indicating the magnetic field values for which the CPR and $E_J(\phi)$ are plotted; CPRs for the three values of magnetic fields; Josephson energies as functions of the phase difference.
The dot in the bottom row $E_J$ indicates the energy minimum and the ground state phase difference.
Note that identical model parameters are used in columns (a),(b) and (f),(g) respectively, but different consecutive junctions states of interest are highlighted in the individual columns. 
\label{fig:CPR}}
\end{figure}

To further support the claims of the previous section and to discuss the role of the ground state phase, we plot the evolution of the critical current and the ground state phase difference with magnetic field, and show the current-phase relations and Josephson energies characteristic for each junction state in Fig.~\ref{fig:CPR}.
$0-\pi$ transitions happen in the absence of spin-orbit interaction (Fig.~\ref{fig:CPR}(a)-(e)). 
In the presence of spin-orbit and disorder, due to breaking of the spatial symmetry the ground state phase can obtain any single value $\varphi_0$ (a so-called $\varphi_0$-junction) near the crossover between $0$ and $\pi$ states of the junction ((Fig.~\ref{fig:CPR}(f)-(g)).
Note that without disorder, the spatial mirror symmetry with respect to the middle of the system forces all CPRs $I(\phi)$ to be odd functions and all $E_J(\phi)$ to be even functions of $\phi$.
When spatial mirror symmetry holds, the junction's Josephson energy can still have a double minimum at $\pm\varphi$ (a so called $\varphi$-junction), thus $E_J(\phi)$ taking a Mexican hat type shape (green curves in (Fig.~\ref{fig:CPR}(b)-(c)).
However, because of this restriction imposed by spatial mirror symmetry, $\varphi$-junctions are rare and most junctions are either 0 or $\pi$-junctions.
Contrarily, including disorder breaks this symmetry leading to commonly occurring $\varphi_0$-junctions.

\section{Effect of disorder}

\begin{figure}[!h]
\centering
\includegraphics[width=0.75\textwidth]{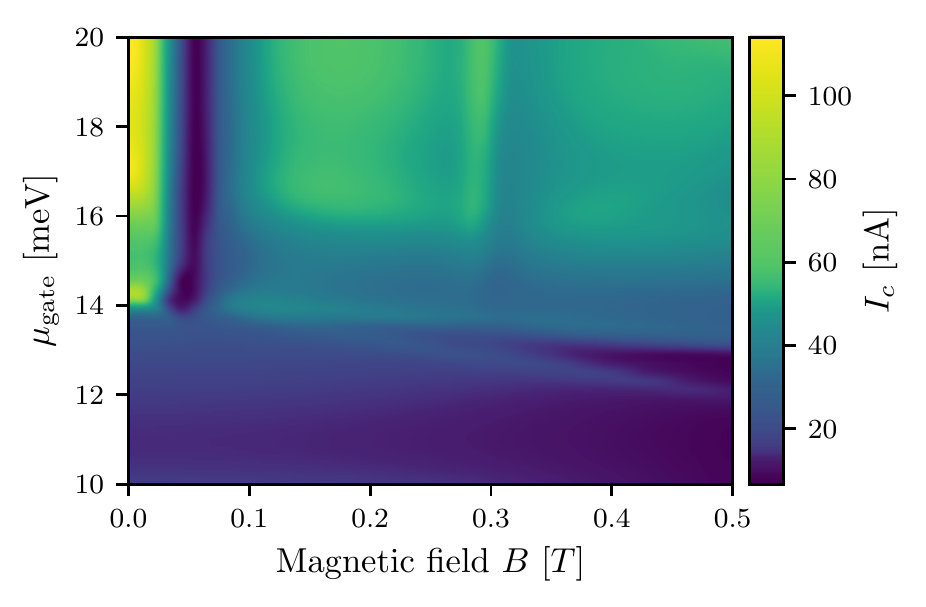}
\caption{Critical current as a function of the magnetic field and the gate voltage. The simulation parameters are identical to the ones used in Fig.~5 in the main text, but we set disorder to zero.\label{fig:gate_dependence}}
\end{figure}

Here we prove the essential effect of disorder on the supercurrent dependence on gate voltage.
We see the effect of disorder on $I_c(V_\textrm{gate})$ by comparing Fig.~5(b) of the main text and Fig.~\ref{fig:gate_dependence}, where we have switched off disorder.
In the clean case, where the main effect of the gate voltage on the supercurrent is via the gradual suppression of transmission through the nanowire, we observe that varying the gate voltage barely causes fluctuations of the supercurrent, even at finite magnetic field.
In the disordered case, changing the gate voltage effectively changes the realization of disorder in the region of the wire above the gate, thus causing supercurrent fluctuations. With the increased disorder, the dwell time in the gated region of the nanowire is increased, so the gate voltage dependence increases with reduced mean free path.
We found that no disorder and disorder with mean free path greater than the system size cannot explain the observed dependence of the critical current on magnetic field and gate voltage.

\section{Rotating magnetic field}
\begin{figure}
\centering
\includegraphics[width=0.75\textwidth]{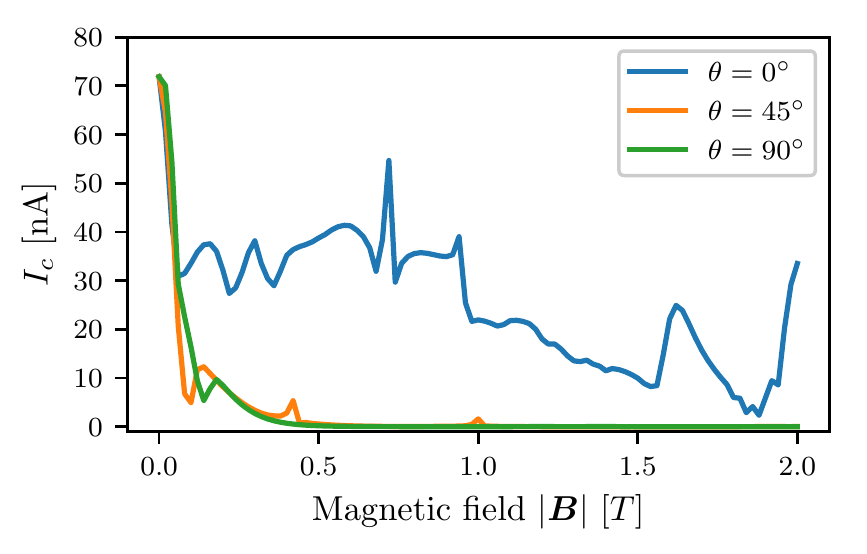}
\caption{
Supercurrent as a function of magnetic field for different directions of the applied field.
The angle is measured with respect to the wire axis and is rotated in the plane $\hat{x}$ parallel to the wire and $\hat{y}$ perpendicular to the wire and parallel to the substrate.
We use the same parameters as in Fig.~5 of the main text\label{fig:rotation_of_field}.
Besides the field purely along $\hat{y}$ the fluctuations pattern is qualitatively similar in all the directions.}
\end{figure}

Here we model the supercurrent fluctuations for different directions of the magnetic field, from parallel to the wire to perpendicular to it. 
The results of the modeling are in Fig.~\ref{fig:rotation_of_field}.
We see that for all directions of the field, besides one parallel to the wire, the fluctuation pattern is basically the same.
This is in accordance with the experimental observations of Fig.~\ref{fig:angle_dependence}. 

\end{document}